\documentclass{emulateapj}
\usepackage{amssymb}
\usepackage{amsmath}
\usepackage{epstopdf}
\usepackage{natbib}
\usepackage[colorlinks, citecolor=blue]{hyperref}
\usepackage[all]{hypcap}

\begin{document}

\title{Modeling The Most Luminous Supernova Associated with a Gamma-Ray
Burst, SN~2011kl}
\author{Shan-Qin Wang\altaffilmark{1,2,3}, Zach Cano\altaffilmark{4},
Ling-Jun Wang\altaffilmark{5}, WeiKang Zheng\altaffilmark{3}, Zi-Gao Dai\altaffilmark{1,2}, \\
Alexei V. Filippenko\altaffilmark{3,6}, and Liang-Duan Liu\altaffilmark{1,2}}

\begin{abstract}
We study the most luminous known supernova (SN) associated with a gamma-ray
burst (GRB), SN~2011kl.
The photospheric velocity of SN~2011kl around peak brightness is
$21,000\pm7,000$ km s$^{-1}$. Owing to different assumptions related to the
light-curve (LC) evolution (broken or unbroken power-law function) of the
optical afterglow of GRB~111209A, different techniques for the LC
decomposition, and different methods (with or without a near-infrared
contribution), three groups derived three different bolometric LCs for
SN~2011kl. Previous studies have shown that the LCs without an early-time
excess preferred a magnetar model, a magnetar+$^{56}$Ni model, or a white
dwarf tidal disruption event model rather than the radioactive heating
model. On the other hand, the LC shows an early-time excess and dip that
cannot be reproduced by the aforementioned models, and hence the
blue-supergiant model was proposed to explain it. 
Here, we reinvestigate the energy sources powering SN~2011kl. We find that
the two LCs without the early-time excess of SN~2011kl can be explained by
the magnetar+$^{56}$Ni model, and the LC showing the early excess can be
explained by the magnetar+$^{56}$Ni model taking into account the cooling
emission from the shock-heated envelope of the SN progenitor, demonstrating
that this SN might primarily be powered by a nascent magnetar.
\end{abstract}


\affil{\altaffilmark{1}School of Astronomy and Space Science, Nanjing
University, Nanjing 210093, China; dzg@nju.edu.cn}
\affil{\altaffilmark{2}Key Laboratory of Modern Astronomy and Astrophysics (Nanjing
University), Ministry of Education, China}
\affil{\altaffilmark{3}Department
of Astronomy, University of California, Berkeley, CA 94720-3411, USA}
\affil{\altaffilmark{4}Instituto de Astrof\'isica de Andaluc\'ia (IAA-CSIC),
Glorieta de la Astronom\'ia s/n, E-18008, Granada, Spain}
\affil{\altaffilmark{5}Astroparticle Physics,
Institute of High Energy Physics,
Chinese Academy of Sciences, Beijing 100049, China}
\affil{\altaffilmark{6}Miller Senior Fellow, Miller Institute for Basic
Research in Science, University of California, Berkeley, CA 94720, USA}

\section{Introduction}

\label{sec:Intro}

The optical spectra of Type Ic supernovae (SNe~Ic) lack hydrogen and helium
absorption lines (see, e.g., \citealt{Fil1997,Mat2011} for reviews). A small
fraction of SNe~Ic have spectra with broad absorption troughs indicating
large photospheric velocities and were dubbed ``broad-lined SNe~Ic
(SNe~Ic-BL)" \citep{Woo2006}. Some SNe~Ic-BL have very large kinetic
energies ($\gtrsim 10^{52}$ erg; e.g.,
\citealt{Iwa1998,Maz2000,Maz2002,Maz2005,Val2008}) and were coined
``hypernovae" \citep{Iwa1998}. Hypernovae are usually rather luminous (with
peak luminosities $\sim 10^{43}$ erg s$^{-1}$), indicating that they must
have synthesized $\sim 0.5$ M$_\odot$ of $^{56}$Ni if they can be explained
by the radioactive heating model \citep{Arn1982,Cap1997,Val2008,Cha2012}.
Moreover, some SNe~Ic-BL are associated with long-duration gamma-ray bursts
(GRBs; e.g., \citealt{Gal1998,Sta2003,Hjo2003}) and have been called
``GRB-SNe" (see \citealt{Woo2006,Hjo2012,Cano2017}, and references therein).
While most GRB-SNe are Type Ic-BL, not all SNe~Ic-BL are associated with
GRBs. The most luminous SNe~Ic are Type Ic superluminous supernovae (SLSNe;
\citealt{Qui2011,Gal2012}); their peak luminosities are $\gtrsim
7\times10^{43}$ erg s$^{-1}$ \citep{Gal2012}, tens of times greater than
those of ordinary SNe.

In this paper, we study a very luminous SN~Ic, SN~2011kl, which is
associated with an ultralong ($T_{90}\approx 10^4$~s) GRB~111209A
(see, e.g., \citealt{Pal2011,Gol2011,Gen2013,Str2003,Gomp2017}) at
redshift $z=0.677$ \citep{Vre2011}. By using WMAP $\Lambda$CDM cosmology
\citep{Spe2003} and adopting H$_0 =71$~km~s$^{-1}$~Mpc$^{-1}$, $\Omega_M$ =
0.27, and $\Omega_\Lambda$= 0.73, \citet{Kann2016} derived a luminosity
distance ($D_L$) of 4076.5~Mpc for GRB~111209A/SN~2011kl. With a peak
luminosity $\sim 3.63^{+0.17}_{-0.16} \times 10^{43}$ erg s$^{-1}$
\citep{Kann2016}\footnote{\citet{Gre2015} found a slightly lower peak
luminosity ($\sim 2.8^{+1.2}_{-1.0}\times 10^{43}$ erg s$^{-1}$) since they
ignored the near-infrared contribution.}, SN~2011kl is the most luminous
GRB-SN yet detected and is significantly more luminous than all other
GRB-SNe, whose average peak luminosity is $1 \times 10^{43}$ erg s$^{-1}$
with a standard deviation of $0.4 \times 10^{43}$ erg s$^{-1}$
\citep{Cano2017}. The time of peak brightness of SN~2011kl is $\sim 14$
days, slightly larger than the average (13.0 days with a standard deviation
of 2.7 days; \citealt{Cano2017}) of other GRB-SNe.

The peak luminosity of SN~2011kl is comparable to that of some
not-quite-superluminous ``SLSNe" (e.g.,
PTF11rks, PTF10hgi, iPTF15esb, iPTF16bad, and PS15br) and commensurate with
the SN-SLSN gap transients observed by \citet{Arc2016}; moreover,
\citet{Liu2017c} pointed out that the spectrum of SN~2011kl is very
different from those of GRB-SNe, but consistent with that of SLSNe. So, the
energy sources powering its LC might be similar to those powering the
SLSN~LCs instead of the ordinary SNe~Ic. The most popular prevailing models
for explaining SLSNe are the magnetar model
\citep{Mae2007,Kas2010,Woos2010,Cha2012,Cha2013,Inse2013,Chen2015,Wang2015a,
Wang2016b,Dai2016,Mor2017}
and the ejecta plus circumstellar medium (CSM) interaction model
\citep{Che1982,Che1994,Chu1994,Cha2012,Cha2013,Liu2017a}.
\footnote{Almost all SLSNe (except maybe pair-instability SLSNe ---
e.g., SN~2007bi \citealt{Gal2009};
but see \citealt{Des2012} and \citealt{Nich2013}) cannot be explained by the
radioactive heating model \citep{Qui2011,Inse2013,Nich2013,Nich2014}. Some
SLSNe~I exhibiting late-time rebrightening and H$\alpha $ emission lines
\citep{Yan2015,Yan2017} might be powered by a hybrid model containing
contributions from magnetars and ejecta-CSM interaction \citep{Wang2016c}.}

There have been several papers modeling the LCs of SN~2011kl. The LC
obtained by \citet{Gre2015} (hereafter the G15~LC) cannot be explained by
the $^{56}$Ni model but can be by the magnetar model \citep{Gre2015,Cano2016}
\footnote{\citet{Cano2016} showed that GRB~111209A and SN~2011kl can be
explained purely by the magnetar model, both for the afterglow and
SN-powered phases.}, the magnetar+$^{56}$Ni model \citep{Met2015,Bers2016},
the white dwarf tidal disruption event (WD~TDE) model \citep{Ioka2016},
and the collapsar model \citep{Gao2016} involving a stellar-mass
black hole and a fallback accretion disk \citep{Woos1993,Mac1999}. By
subtracting the contribution from the ultraviolet (UV)/optical/near-infrared
(NIR) afterglow \citep{Kann2017}
of GRB~111209A, \citet{Kann2016} obtained a new bolometric LC
(hereafter the K16~LC) and modeled it with the radioactive heating model,
finding $2.27 \pm 0.64$ M$_{\odot }$ of $^{56}$Ni and $6.79_{-2.84}^{+3.67}$ M$_{\odot}$
of ejecta, suggesting that this model is disfavored. The LC derived by
\citet{Ioka2016} (hereafter the I16~LC) has an early-time excess and a dip
\footnote{This early-time excess arose from the assumption that a
single power law
could describe the evolution of the optical LC during the afterglow phase. A
broken power law showed no such excess (e.g., \citealt{Gre2015,Kann2016}),
so this excess is model-dependent. Moreover, a single power law was used by
\citet{Cano2016}, and no such excess was observed, making it inconclusive.}
which cannot be reproduced by the models mentioned above, but it can be fit
by the blue supergiant (BSG) model with explosive energy injection
(``the BSG model"; \citealt{Kash2013,Nakau2013,Ioka2016}).
However, it is premature to exclude
the magnetar model in explaining the I16~LC since the cooling emission from
the shock-heated extended envelope \citep{Piro2015} or the emission from
magnetar shock breakout \citep{Kas2016} can reproduce the early-time excess.

Thus, the issue of the energy source of SN~2011kl deserves additional
attention. In this paper, we reinvestigate the possible mechanisms powering
the LC of SN~2011kl and discuss their implications. In Section \ref{sec:fit}
, we use the radioactive heating model, the magnetar model, the
magnetar+$^{56}$Ni model, and the
magnetar+$^{56}$Ni+cooling model to fit the LCs of
SN~2011kl. Our discussion and conclusions are presented in Sections
\ref{sec:dis} and \ref{sec:con}, respectively.

\section{Modeling the Light Curves of SN~2011kl}

\label{sec:fit}

\begin{figure}[tbph]
\begin{center}
\includegraphics[width=1.00\textwidth,angle=0]{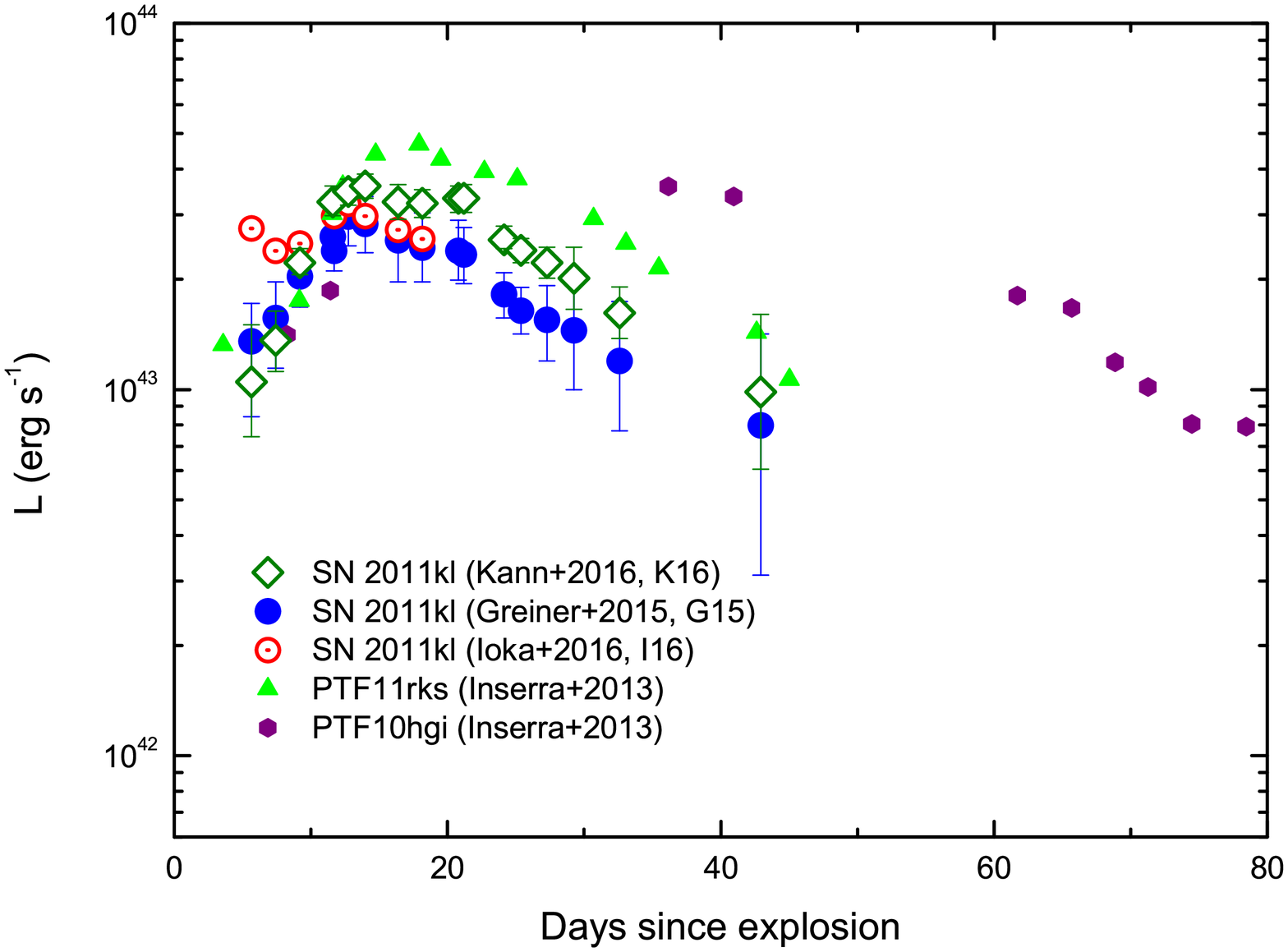}
\end{center}
\caption{LCs of SN~2011kl. The data are taken from \citet{Gre2015},
\citet{Ioka2016}, and \citet{Kann2016}. For comparison, the early-time
($t\leq80$ days) LCs of SLSNe PTF11rks and PTF10hgi \citep{Inse2013} are also
shown. The horizontal axis represents the time since the explosion in the
rest frame.}
\label{fig:LCs}
\end{figure}

Extracting the LC of a SN associated with a GRB is difficult since this
process requires a decomposition of the observations to account for the
contributions from the optical afterglow, the SN, and the underlying host
galaxy (e.g., \citealt{Zeh2004}); moreover, it depends on the assumptions
adopted for modeling the GRB optical afterglow. By assuming that the LC of
the optical afterglow of GRB~111209A can be described by a broken power-law
function and subtracting the contributions from the optical afterglow and
the host galaxy, \citet{Gre2015} obtained the LC of SN~2011kl (the G15~LC).
The G15~LC did not include the NIR contribution and is a
pseudobolometric LC. By adding the NIR contribution, \citet{Kann2016}
obtained a genuine bolometric LC (the K16~LC)\footnote{The
K16~LC must be brighter than the pseudobolometric LC (the G15~LC)
without the NIR contribution. However, the median values of the luminosities
of the first two points of the K16~LC are dimmer than that of the G15~LC.
This fact is puzzling, perhaps indicating that the uncertainties are
rather large.}. However, based on the assumption that the optical afterglow
of GRB~111209A can also be described by a single (unbroken) power-law
function, \citet{Ioka2016} derived another LC (the I16~LC) for SN~2011kl.
This LC has an early-time excess and a dip. Hence, SN~2011kl has three
different LCs (see Figure \ref{fig:LCs}) that must be modeled.

The LCs of SLSNe PTF11rks and PTF10hgi \citep{Inse2013} are also plotted in
Figure \ref{fig:LCs}. By comparing the K16~LC of GRB-SN~2011kl, whose peak
luminosity $L_{\mathrm{peak,bol}}$ is $\sim 3.6 \times 10^{43}$ erg s$^{-1}$,
to the LCs of SLSNe PTF11rks and PTF10hgi, we find that the peak of the LC
of SN~2011kl is as luminous as that of these two SLSNe. Furthermore, we
point out that other SLSNe also have similar values of $L_{\mathrm{peak,bol}}$
 --- e.g., iPTF15esb ($L_{\mathrm{peak,bol}}\approx 4\times 10^{43}$
erg s$^{-1}$; \citealt{Yan2017}), iPTF16bad ($L_{\mathrm{peak,bol}}\approx
4\times 10^{43}$ erg s$^{-1}$; \citealt{Yan2017}), and PS15br
($L_{\mathrm{peak,bol}} \approx 4.15\times 10^{43}$ erg s$^{-1}$;
\citealt{Inse2016}).\footnote{SN~2011kl,
along with PTF11rks, PTF10hgi, iPTF15esb, iPTF16bad, and PS15br,
are dimmer than the strict threshold for SLSNe ($L_{\mathrm{peak}}>7\times
10^{43}$ erg s$^{-1}$ and the peak absolute magnitudes
$M_{\mathrm{peak}}<-21$ mag \emph{in any band}; \citealt{Gal2012}).}

In this section, we use some models to fit these LCs. For the G15 and
K16 LCs, we adopt the Markov Chain Monte Carlo (MCMC) method to minimize
the values of $\chi^2$/dof and get the best-fit parameters and the range.
For the I16 LC, we don't use the MCMC method because its data have no
known error bars.

\subsection{The Radioactive Heating Model}

Based on their derived LC (the G15~LC), \citet{Gre2015} fit the $^{56}$Ni
model and found that the required mass of $^{56}$Ni is $\sim 1$~M$_\odot$ if
the LC was powered by $^{56}$Ni decay. \citet{Kann2016} suggested that their
bolometric LC (the K16~LC), which included the NIR contribution, needed
$2.27 \pm 0.64$ M$_\odot$ of $^{56}$Ni to reproduce its luminosity if the LC
was powered by radioactive heating.

Here we employ the semi-analytic $^{56}$Ni model
\citep{Arn1982,Cap1997,Val2008,Cha2012} to repeat the fits for the G15~LC,
the K16~LC and I16LC. The free parameters of the $^{56}$Ni model are the
optical opacity $\kappa$, the ejecta mass $M_{\mathrm{ej}}$, the $^{56}$Ni
mass $M_{\mathrm{Ni}}$, the initial scale velocity of the
ejecta\footnote{The
scale velocity of the ejecta $v_{\mathrm{sc}}$ is the velocity of the
surface of the ejecta. The radius of the ejecta can be calculated from $R(t)
= R(0)+v_{\mathrm{sc}}t$ \citep{Arn1982}.} $v_{\mathrm{sc0}}$, and the
gamma-ray opacity of $^{56}$Ni-cascade-decay photons
$\kappa_{\gamma,\mathrm{Ni}}$.
The value of $\kappa$ is rather uncertain and has been
assumed to be 0.06 cm$^2$~g$^{-1}$ (e.g., \citealt{Val2011,Lym2016}), 0.07
cm$^2$ g$^{-1}$ (e.g., \citealt{Chu2000,Tad2015,Wang2015b}), $0.07\pm0.01$
cm$^2$~g$^{-1}$ (e.g., \citealt{Gre2015,Kann2016}), 0.08 cm$^2$~g$^{-1}$
(e.g., \citealt{Arn1982,Maz2000}), 0.10 cm$^2$~g$^{-1}$ (e.g.,
\citealt{Inse2013,Whe2015}), and 0.20 cm$^2$~g$^{-1}$ (e.g.,
\citealt{Kas2010,Nich2014,Arc2016}). Assuming that the dominant opacity source
is Thomson electron scattering and the temperature of the SN ejecta consisting
of carbon and oxygen is not very high ($\lesssim 10,000$ K), we adopted 0.07~
cm$^2$~g$^{-1}$ to be the value of $\kappa$. The value of $\kappa_{\gamma,\mathrm{Ni}}$
is fixed to be 0.027 cm$^{2}$ g$^{-1}$ (e.g.,
\citealt{Cap1997,Maz2000,Mae2003,Whe2015}).
 The photospheric velocity of SN~2011kl inferred from the spectra is $\sim
21,000\pm7,000$ km s$^{-1}$ \citep{Kann2016}.
We assume that the moment of explosion ($t_\mathrm{expl}$) for SN
2011kl is equal to that of GRB~111209A.

The LCs reproduced by the radioactive heating models A1, A2, and A3 are
shown in Figure \ref{fig:2011kl}, and the best-fit parameters are listed
in Table \ref{tab:para}. While the best-fit model parameters for the
G15~LC are approximately equal to that derived by \citet{Gre2015}, our
inferred $^{56}$Ni mass ($1.42_{-0.04}^{+0.04}$ M$_\odot$) for the
K16~LC is smaller than the value derived by \citet{Kann2016} ($2.27\pm0.64$
M$_\odot$). Moreover, the value of the ejecta mass
($4.57_{-1.03}^{+0.80}$ M$_\odot$) is also smaller than that
derived by \citet{Kann2016} ($6.79_{-2.84}^{+3.67}$ M$_\odot$).

Another method for estimating the value of $^{56}$Ni is to use the ``Arnett
law" \citep{Arn1982}, which says that the $^{56}$Ni energy input LC
intersects the peak of the SN~LC. According to the Arnett law
and Eq. 19 of \citep{Nady1994}, the mass of
$^{56}$Ni is \footnote{Eq. 19
of \citet{Nady1994} is valid only in the case of full gamma-ray
trapping. For partial gamma-ray trapping ($\kappa_{\gamma,\mathrm{Ni}}$
= 0.027 cm$^2$ g$^-1$), however, the inferred $^{56}$Ni mass is equal to
that corresponding to the case of full gamma-ray trapping, because the gamma-ray
leakage doesn't influence the luminosity before and around the LC peak.}
\begin{eqnarray}
M_{\mathrm{Ni}}=({L_{\mathrm{peak}}}/{10^{43}~\mathrm{erg}~\mathrm{s}^{-1}})
{[6.45~e^{-t_{\mathrm{rise}}/8.8~\mathrm{days}}}  \notag \\
{+1.45~e^{-t_{\mathrm{rise}}/111.3~\mathrm{days}}]}^{-1}~\mathrm{M}_\odot,
\end{eqnarray}
\noindent
since the value of $t_{\mathrm{rise}}$ of SN~2011kl is $\sim 14$ days and
the peak luminosity is $3.63^{+0.17}_{-0.16}\times 10^{43}$~erg~s$^{-1}$
(for the K16~LC) or $2.8^{+1.2}_{-1.0}\times 10^{43}$~erg~s$^{-1}$ (for the
G15~LC), the $^{56}$Ni mass is $1.40^{+0.07}_{-0.06}$ M$_\odot$ (for the
K16~LC) or $1.08^{+0.46}_{-0.39}$ M$_\odot$ (for the G15~LC).

\subsection{The Magnetar Model}

\citet{Gre2015} analyzed the spectrum of SN 2011kl and pointed out that the
spectral features imply that the $^{56}$Ni mass must be significantly
smaller than 1 M$_\odot$; they excluded the $^{56}$Ni model for explaining
the LC of SN~2011kl. Alternatively, \citet{Gre2015} employed the magnetar
model developed by \citet{Kas2010} to fit the G15~LC and obtained a rather
good result. But the K16~LC has not yet been fit with the magnetar model.

Here, we use the semianalytic magnetar model developed by \citet{Wang2015a}
and \citet{Wang2016b} that considers the leakage, photospheric recession,
and acceleration effects to fit the K16~LC, and we repeat the fits for the
G15~LC. The free parameters of the magnetar model are $\kappa $,
$M_{\mathrm{ej}}$,
the magnetic strength of the magnetar $B_{p}$, the initial rotation
period of the magnetar $P_{0}$, $v_{\mathrm{sc0}}$, and the gamma-ray
opacity of magnetar photons $\kappa _{\gamma ,\mathrm{mag}}$. The
value of $\kappa _{\gamma ,\mathrm{mag}}$ depends mainly on the energy of
the photons ($E_{\mathrm{photons}}$) emitted from the nascent magnetars,
varying between $\sim 0.01$ and 0.2 cm$^{2}$~g$^{-1}$ for $\gamma$-ray
photons ($E_{\gamma }\gtrsim 10^{6}$~eV)
and between $\sim 0.2$ and $10^{4}$~cm$^{2}$~g$^{-1}$
for X-ray photons ($10^{2}$~eV $\lesssim E_{\mathrm{X}}\lesssim
10^{6}$~eV); see Fig. 8 of \citet{Kot2013}.

The LCs reproduced by the magnetar models (B1, B2, and B3) are shown in
Figure \ref{fig:2011kl} and the best-fit parameters are listed in Table
\ref{tab:para}. We find that the K16~LC can be powered by a magnetar having
$B_{p} = 5.99_{-5.55}^{+1.75} \times10^{14}$~G and $P_{0} =
10.83_{-6.94}^{+0.49}$~ms, while the G15~LC can be powered by a magnetar
with $B_{p} = 9.72_{-4.13}^{+3.23} \times10^{14}$~G and $P_{0} =
12.07_{-2.52}^{+1.15}$~ms (the values of $B_{p}$ and $P_{0}$ derived by
\citet{Gre2015} are $\sim$ (6--9) $\times10^{14}$~G and $\sim 13.4$~ms,
respectively).

\subsection{The Magnetar Plus $^{56}$Ni Model}

The contribution of $^{56}$Ni cannot be ignored when modeling some luminous
SNe~Ic \citep{Wang2015b,Met2015,Bers2016}. \citet{Wang2015b} instead used a
hybrid model containing contributions from $^{56}$Ni and a magnetar to fit
three luminous SNe~Ic-BL whose peak luminosities are approximately equal to
that of SN~2011kl. \citet{Met2015} and \citet{Bers2016} used the same model
to fit SN~2011kl, suggesting that 0.2 M$_\odot$ of $^{56}$Ni must be added
so that a better fit can be obtained. Therefore, we used the hybrid model
(the $^{56}$Ni+magnetar model) to fit the K16~LC and the G15~LC. The free
parameters of this model are $\kappa$, $M_{\mathrm{ej}}$, $M_{\mathrm{Ni}}$,
$B_{p}$, $P_{0}$, $v_{\mathrm{sc0}}$, $\kappa_{\gamma,\mathrm{Ni}}$, and
$\kappa_{\gamma,\mathrm{mag}}$.

Since an energetic SN explosion can synthesize $\sim 0.2$ M$_\odot$
of $^{56}$Ni \citep{Nom2013,Mae2009}, we assumed that the range of $^{56}$Ni
is 0--0.2 M$_\odot$ (for the G15 and K16 LCs with error bars, which allow us to
use the MCMC method and determine the best-fit parameters) or 0.2 M$_\odot$
(for the I16 LC without error bars). The LCs reproduced by the $^{56}$
Ni+magnetar model (C1, C2, and C3) are shown in Figure \ref{fig:2011kl} and
the best-fit parameters are listed in Table \ref{tab:para}. We find that
this hybrid model can also reproduce good fits, and that almost all
parameters (except for $P_{0}$) adopted by the $^{56}$Ni+magnetar model are
the same as that of the magnetar models. Moreover, the masses of
$^{56}$Ni determined by the MCMC are $0.11_{-0.07}^{+0.06}$ M$_\odot$ (for
the K16 LC) and $0.10_{-0.07}^{+0.06}$ M$_\odot$ (for the G16 LC).

\subsection{The magnetar plus $^{56}$Ni plus cooling model}

\begin{figure}[tbph]
\begin{center}
\includegraphics[width=0.45\textwidth,angle=0]{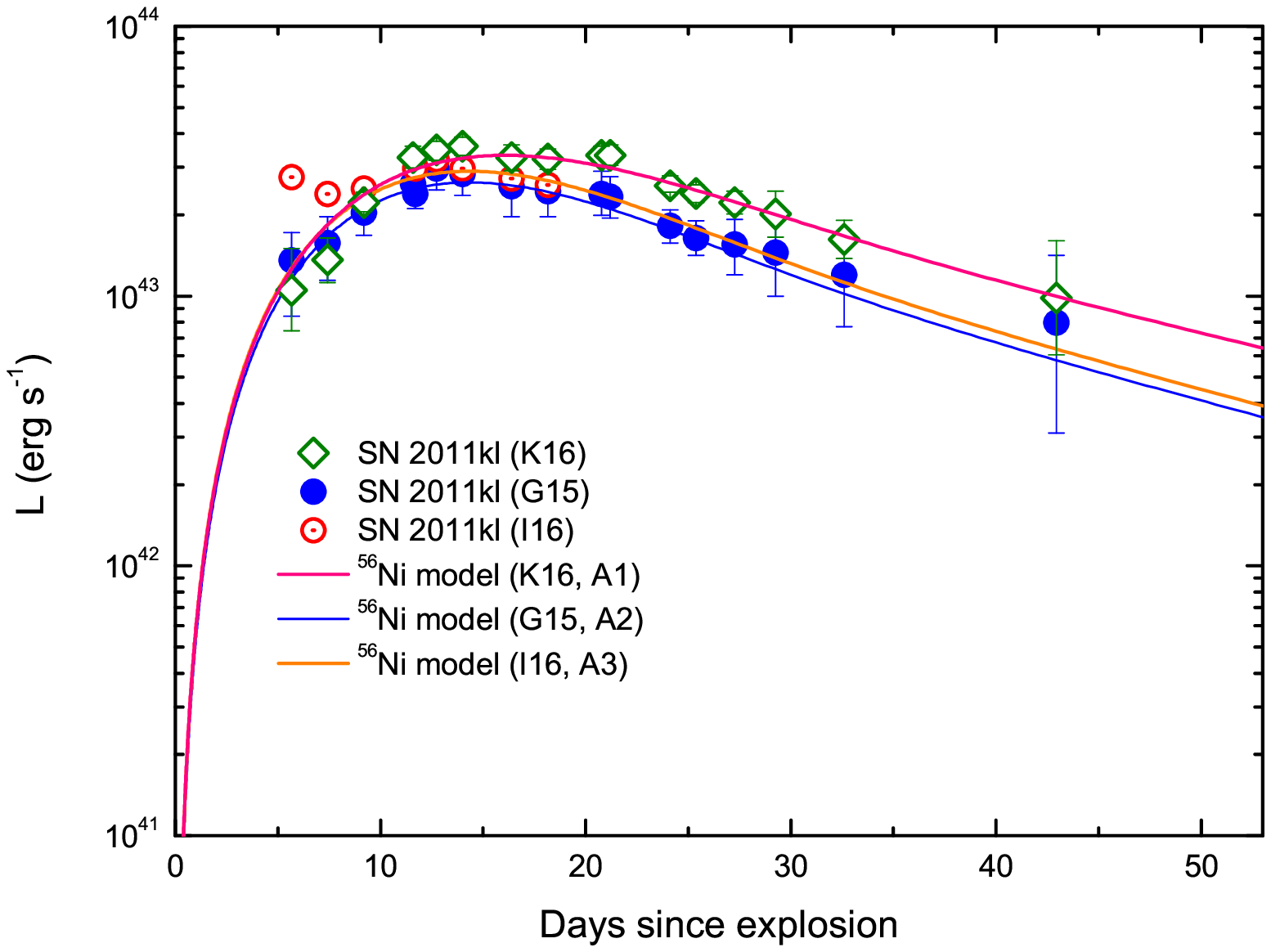}
\includegraphics[width=0.45\textwidth,angle=0]{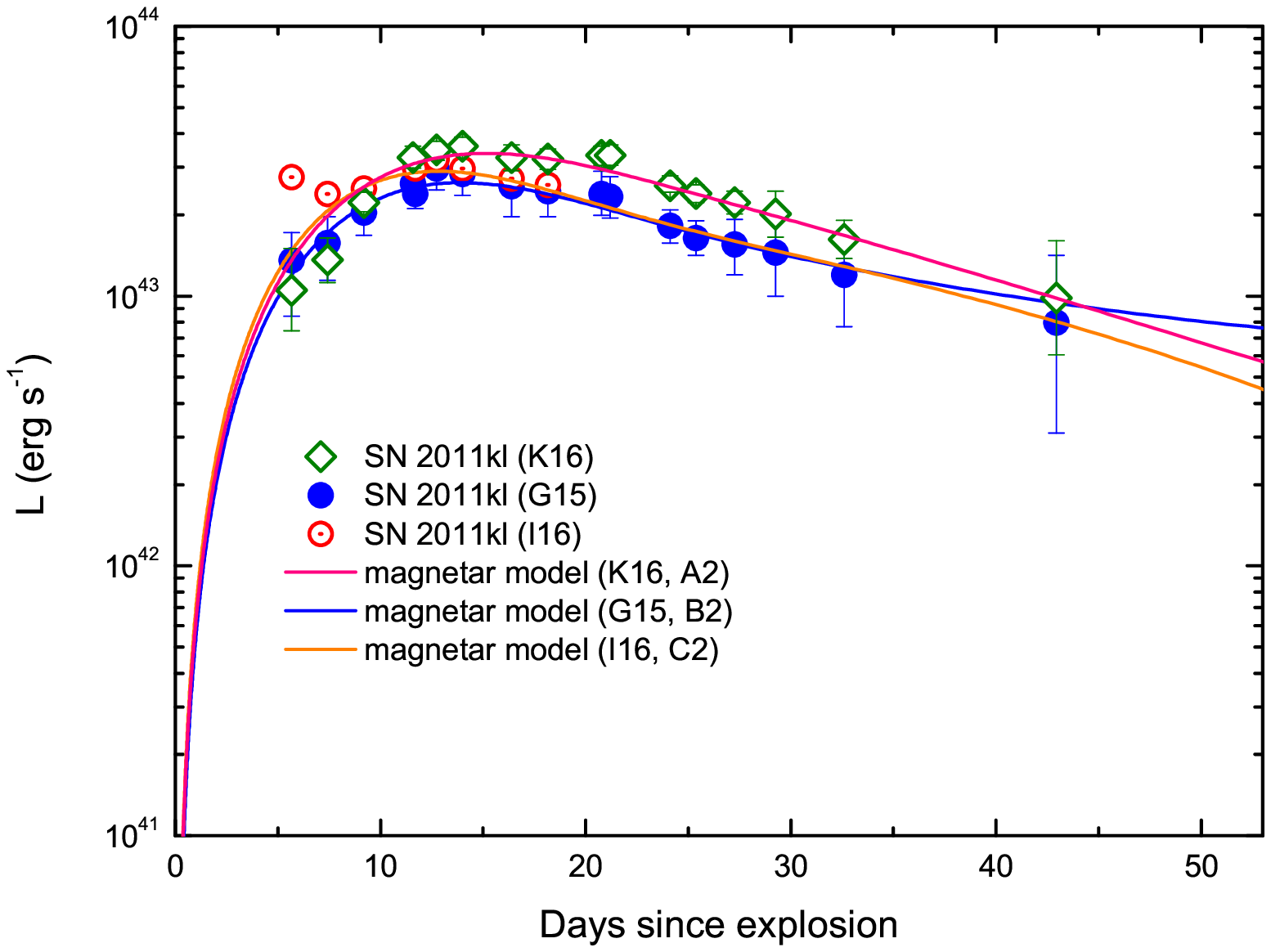}\\[0pt]
\includegraphics[width=0.45\textwidth,angle=0]{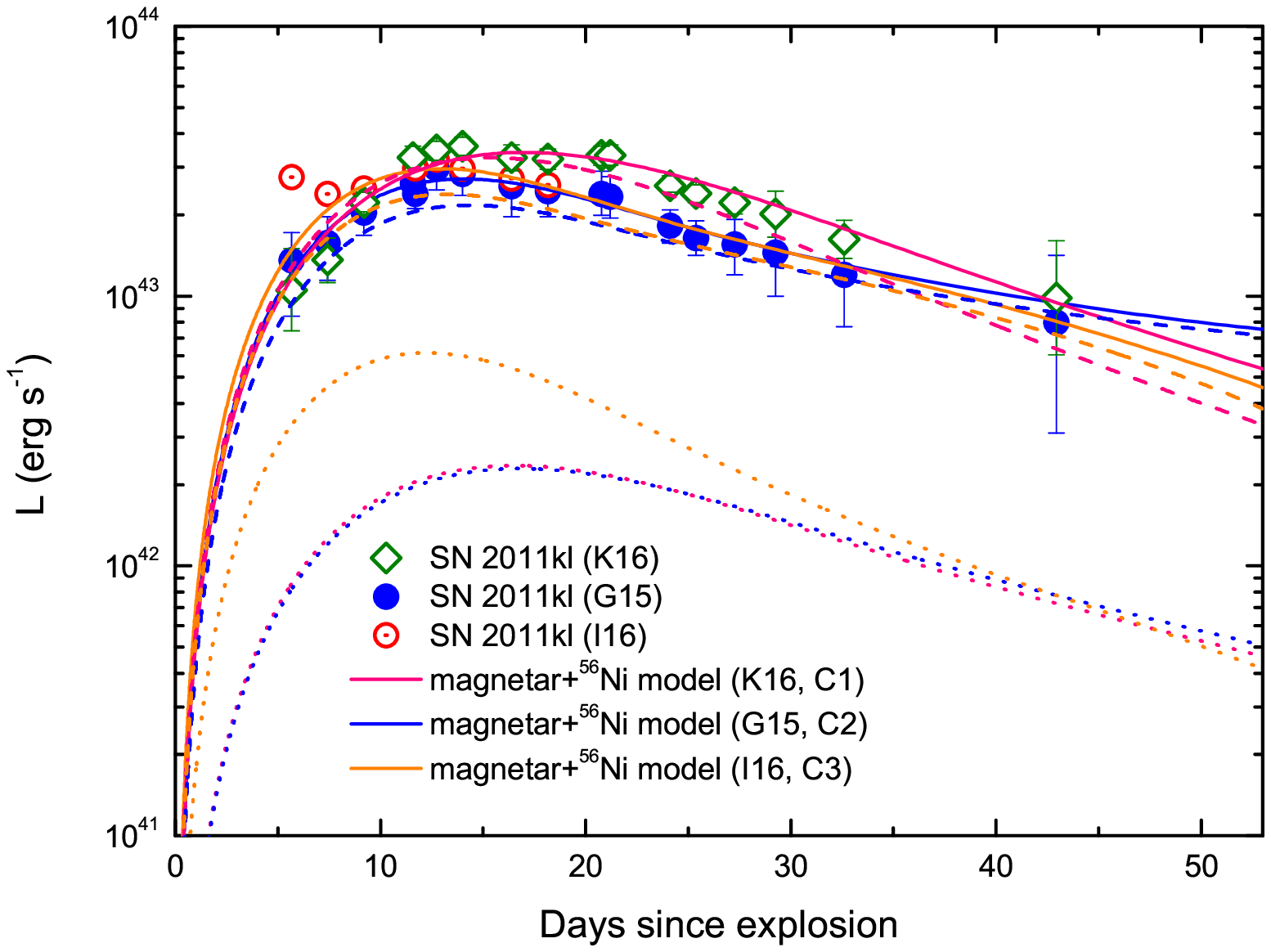}
\includegraphics[width=0.45\textwidth,angle=0]{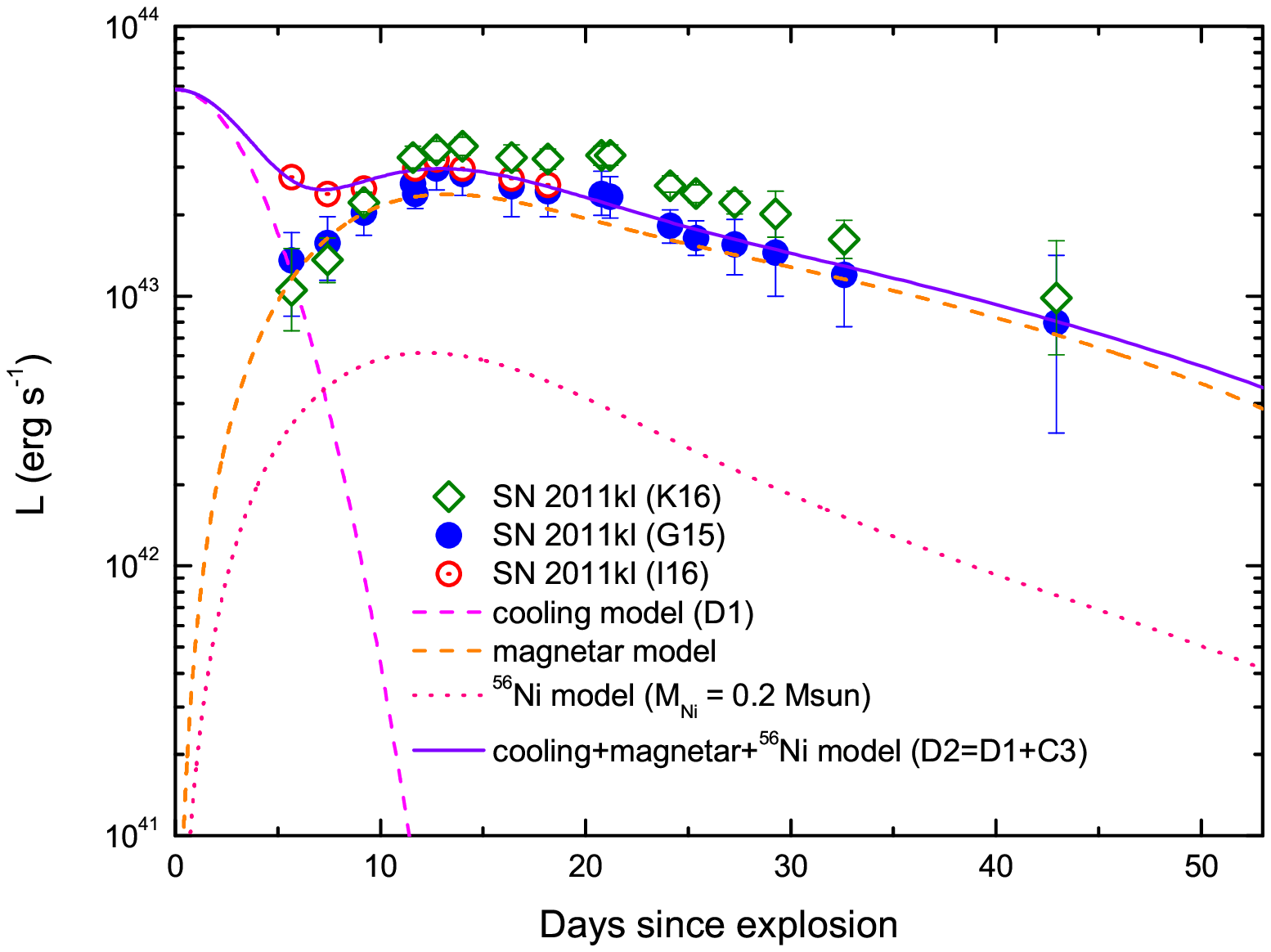}
\end{center}
\caption{The LCs reproduced by different models (A1--A3, B1--B3, C1--C3, and
D1--D4) listed in Table \protect\ref{tab:para}. Data for SN~2011kl are taken
from \citet{Gre2015}, \citet{Ioka2016}, and \citet{Kann2016}. The
thin dotted lines in the lower panels represent the LCs reproduced by 0.1
(the dimmer LCs) or 0.2 M$\odot$ (the brighter LCs) of $^{56}$Ni. The
dashed lines in the lower panels represent the LCs reproduced by the magnetar. The 
horizontal axis represents the time since the explosion in the rest frame.}
\label{fig:2011kl}
\end{figure}

The I16~LC has an early-time excess which cannot be reproduced by the
$^{56}$Ni model,
the magnetar model, or the magnetar+$^{56}$Ni model (A3, B3, and
C3, respectively; see Figure \ref{fig:2011kl}). The WD-TDE model proposed by
\citet{Ioka2016} also cannot explain the I16~LC (see Figure 1 of
\citealt{Ioka2016}). \citet{Ioka2016} suggested that the I16~LC can be
explained by the BSG model.

Here, we use another model to explain the I16~LC. We suggest that the
early-time excess might be due to the cooling emission from the shock-heated
extended envelope \citep{NP2014,Piro2015}, while the main peak of the LC
might be powered by a magnetar or a magnetar plus a moderate amount of
$^{56}$Ni. In this scenario, the progenitor of the SN is surrounded by an
extended, low-mass envelope which is heated by the SN shock and produces a
declining bolometric LC when it cools. At the same time, sources at the
center of the SN release energy and produce a rising LC until it peaks. By
combining the magnetar(+$^{56}$Ni) model developed by \citet{Wang2015b} and
\citet{Wang2016b} and the cooling model developed by \citet{Piro2015}, a LC
with an early-time excess and a dip can be produced. In this model, two
additional parameters must be added: the mass of the extended envelope
($M_{\mathrm{env}}$)
and the radius of the extended envelope ($R_{\mathrm{env}}$).
\footnote{The energy of the SN, $E_{\mathrm{SN}}$, must also be added into the
modeling. However, it can be supposed that
$E_{\mathrm{SN}}=E_{\mathrm{K}}=
0.3\,M_{\mathrm{ej}}v_{\mathrm{sc}}^{2}$, so this parameter is determined
by $M_{\mathrm{ej}}$ and $v_{\mathrm{sc}}$, while $v_{\mathrm{sc}}$ is
determined by $v_{\mathrm{sc0}}$, $B_{p}$, and $P_{0}$ (the acceleration
effect).}

The LCs reproduced by the cooling model
and the magnetar+$^{56}$Ni+cooling model (D1--D3) are shown in Figure
\ref{fig:2011kl} and the corresponding parameters are listed in Table
\ref{tab:para}. We find that 
the magnetar+$^{56}$Ni+cooling models (D2 and D3) can well fit the entire
I16~LC. In the scenario containing the contribution from the cooling
emission, the progenitor of SN~2011kl was supposed to be surrounded by an
extended envelope whose mass and radius are $\sim 0.63$ M$_\odot$ (or $\sim
0.45$ M$_\odot$) and $\sim 51.4$ R$_\odot$ (or $\sim 103$ R$_\odot$),
respectively. Owing to the parameter degeneracy, there must be other choices
of these two parameters.

One advantage of the magnetar-dominated model is that it can explain the
I16~LC if the cooling emission from the extended envelope is added, and it
can explain the K16~LC and the G15~LC if the progenitor was not surrounded
by an extended envelope. Another advantage is that the LC reproduced by the
model with cooling emission can trace the entire LC (including the early
excess, the dip, the peak, and the post-peak) while the LC reproduced by the
BSG model is a smooth, monotonically decreasing LC that is brighter than the
dip and dimmer than the peak. Although the uncertainties of the early-time
data indicate that the LC reproduced by the BSG model might fit the data,
the magnetar+$^{56}$Ni+cooling model can give a better match.

\subsection{Summary}

In summary, the K16, G15, and I16~LCs cannot be powered solely by the
$^{56}$Ni model since (a) $\sim 1.00$--1.42 M$_\odot$
of $^{56}$Ni is inconsistent
with the spectral analysis performed by \citet{Gre2015}, and (b) the ratio
of $^{56}$Ni masses ($\sim$ 1.0--1.4~M$_\odot$) to the ejecta mass ($\sim$
2--5~M$_\odot$) is $\sim 0.3$, which is significantly larger than the upper
limit ($\sim 0.20$; \citealt{Ume2008}). Alternatively, the K16 and G15~LCs
can be explained by the magnetar model and the $^{56}$Ni+magnetar model; the
values of $M_{\mathrm{ej}}$, $B_{p}$, and $P_0$ can be found in
Table \ref{tab:para}.

Owing to the degeneracy of model parameters, the values of these parameters
have many choices that can also give the best-fit LC. For example,
\citet{Bers2016} obtained $P_0 = 3.5$~ms, $B_{p}=1.95\times10^{15}$~G,
$M_{\mathrm{ej}} = 2.5$ M$_\odot$ if $\kappa = 0.2$ cm$^2$ g$^{-1}$, and
$M_{\mathrm{Ni}} = 0.2$ M$_\odot$; \citet{Met2015} derived $P_0 \approx
2$~ms, $B_{p}\approx 4\times10^{14}$~G, $M_{\mathrm{ej}} = 3$
M$_\odot$ if $\kappa = 0.2$ cm$^2$ g$^{-1}$,
and $M_{\mathrm{Ni}} = 0.2$ M$_\odot$; and
\citet{Cano2016} derived $P_0 = 11.5$--13.0~ms, $B_{p}=$ (1.1--1.3)
$\times10^{15}$~G, and $M_{\mathrm{ej}} = 5.2$ M$_\odot$ if $\kappa =
0.07$ cm $^2$ g$^{-1}$.

\section{Discussion}

\label{sec:dis}

\subsection{What Determines the Peak Luminosities of SNe and SLSNe?}

The inferred best-fit values of $P_0$ and $B_p$ of the magnetar
powering the LCs of SN~2011kl are $\sim$ 9--14~ms and $\sim$ (6--14) $\times
10^{14}$~G, respectively. The values of $P_0$ obtained by \citet{Bers2016},
\citet{Met2015}, and \citet{Cano2016} are 3.5~ms, 2~ms, and 11.5--13.0~ms
(respectively), while the respective values of $B_p$ derived by these groups
are $1.95\times10^{15}$~G, $\sim 4\times10^{14}$~G, and (1.1--1.3)
$\times10^{15}$~G. The spin-down timescale of a magnetar ($\tau_p$) is
determined by the values of $P_0$ and $B_p$,
$\tau_p = 1.3\,(B_p/10^{14}~\mathrm{G})^{-2}(P_0/10~\mathrm{ms})^2$~yr
\citep{Kas2010}. We note that
other SLSNe~Ic whose peak luminosities are larger than $\sim 10^{44}$
erg s$^{-1}$ must be powered by magnetars with $P_0 \approx 1$--10~ms
and $10^{13}$--$10^{15}$ (see, e.g.,
\citealt{Inse2013,Nich2014,Liu2017b,Nich2017,Yu2017}
), indicating that the values of $P_0$ and $B_p$ play a crucial role in
determining the peak luminosities of SLSNe. When $P_0 \gtrsim 10$~ms and
$B_p$ is a few $10^{14}$~G, the magnetar-powered SNe are luminous
\citep{Wang2015b}; when $P_0 \lesssim 10$~ms and $B_p$ is a few $10^{16}$~G,
the magnetars can explain the LCs of some SNe~Ic-BL
(\citealt{Wang2016a,Wang2017a,Wang2017b}; see also \citealt{Chen2017}), while
\citet{Cano2016} suggested that most GRB-SNe are powered by radioactive
heating.

\citet{Wang2015b} emphasized the role of $P_0$ while \citet{Yu2017}
highlighted the role of $B_p$. We suggest that the more reasonable scheme
discussing the factors influencing the peak luminosities and the shapes of
magnetar-powered SNe and SLSNe is to take into account both of these
parameters. However, it should be noted that the values of $\tau_m$ (or
$M_{\mathrm{ej}}$) also influence the peak luminosities of magnetar-powered SNe
and SLSNe \citep{Nich2015}. Hence, the luminosities of SNe and SLSNe are
determined by all of these parameters ($P_0$, $B_p$, and $\tau_m$, which is
primarily determined by $M_{\mathrm{ej}}$) and cannot solely be influenced
by any single one. Similarly, the peak luminosities of $^{56}$Ni-powered SNe
and SLSNe are determined by the values of $M_{\mathrm{Ni}}$ and $\tau_{m}$.

\subsection{Correlations Between the Opacity and Other Parameters}

In Section \ref{sec:fit}, we assumed that the value of $\kappa$ is
0.07 cm$^2$ g$^{-1}$. As mentioned above, however, the value of $\kappa$ is
rather uncertain (0.06--0.20 cm$^2$ g$^{-1}$). Therefore, it is necessary to
explore the correlations between the opacity and all other parameters
involved in the modeling.

Assuming $\kappa$ = 0.10 cm$^2$ g$^{-1}$ and 0.20 cm$^2$ g$^{-1}$,
and using the $^{56}$Ni model, the magnetar model, and the magnetar plus
$^{56}$Ni model, we repeated the fits for the K16 LC and the G15 LC,
obtaining the best-fit parameters (see Table \ref{tab:para}).

\begin{figure}[tbph]
\begin{center}
\includegraphics[width=0.45\textwidth,angle=0]{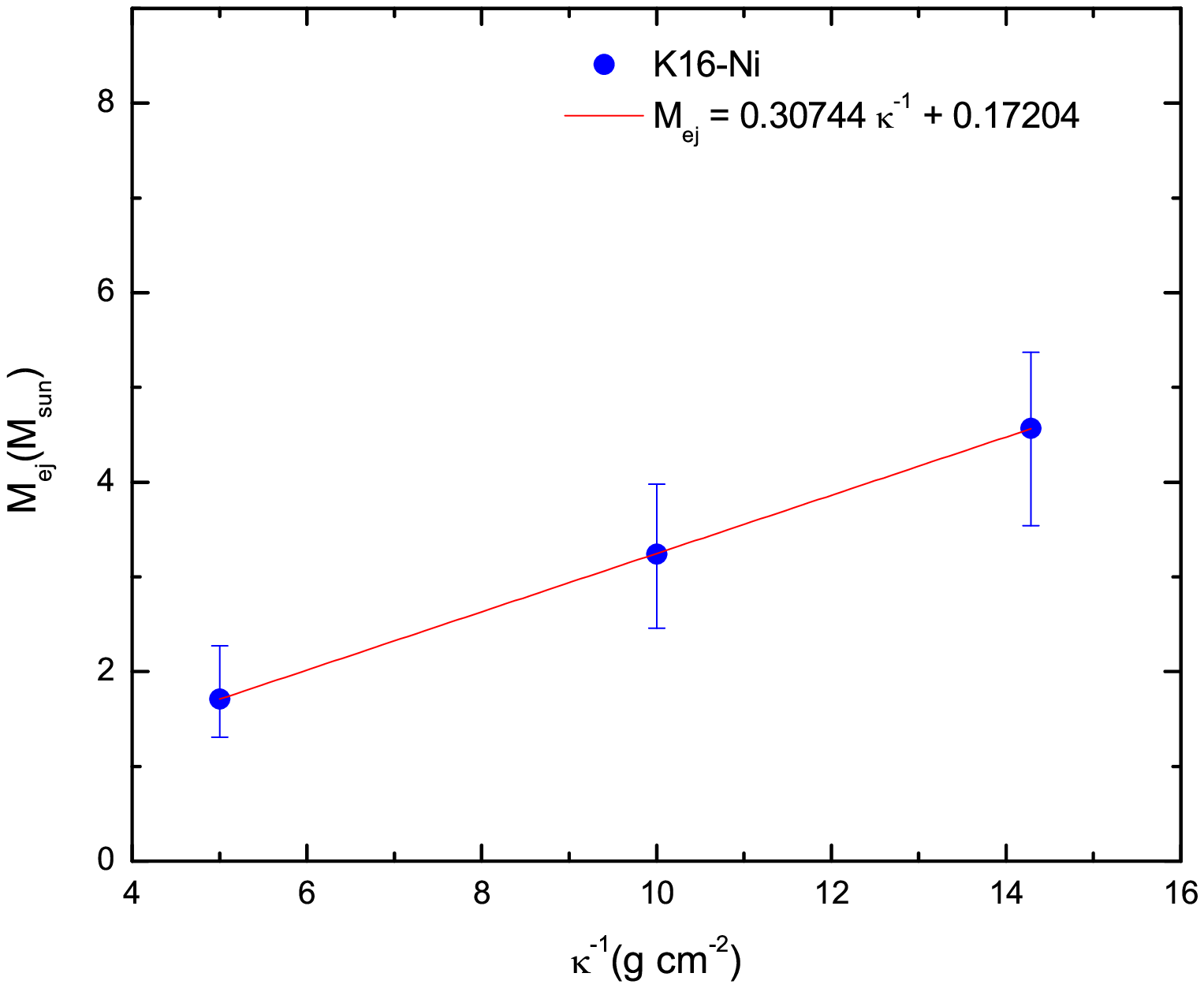}
\includegraphics[width=0.45\textwidth,angle=0]{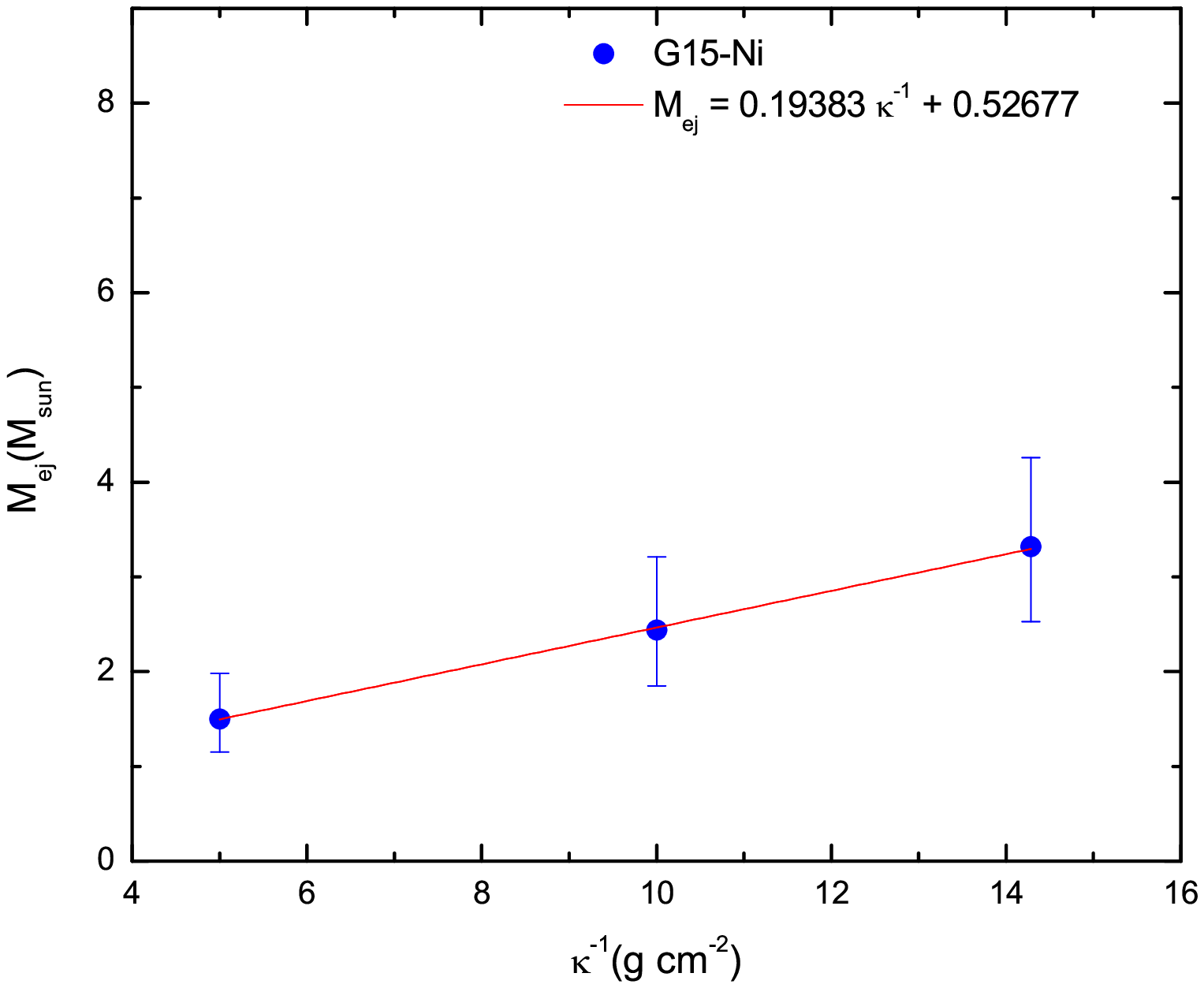}\\[0pt]
\includegraphics[width=0.45\textwidth,angle=0]{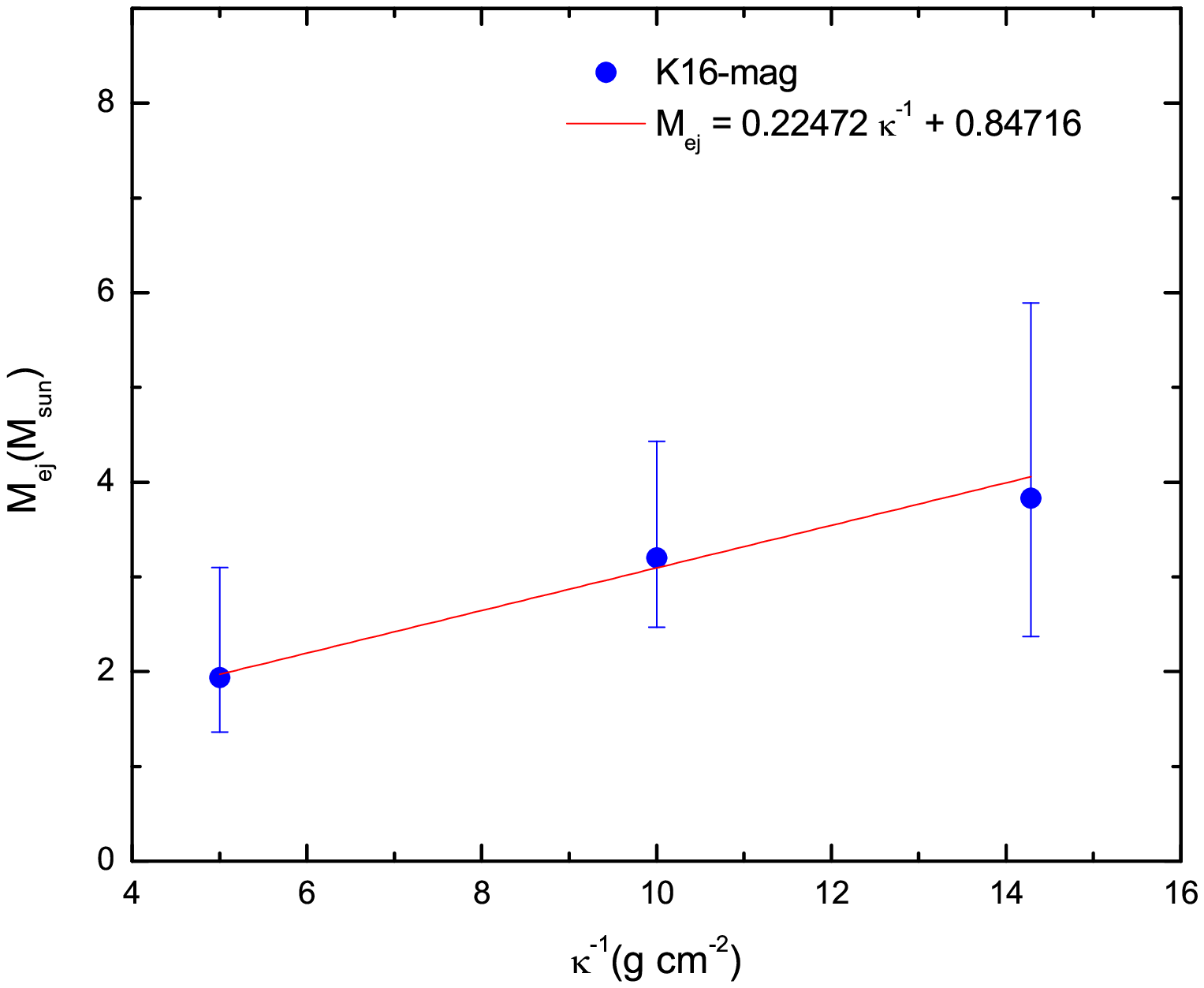}
\includegraphics[width=0.45\textwidth,angle=0]{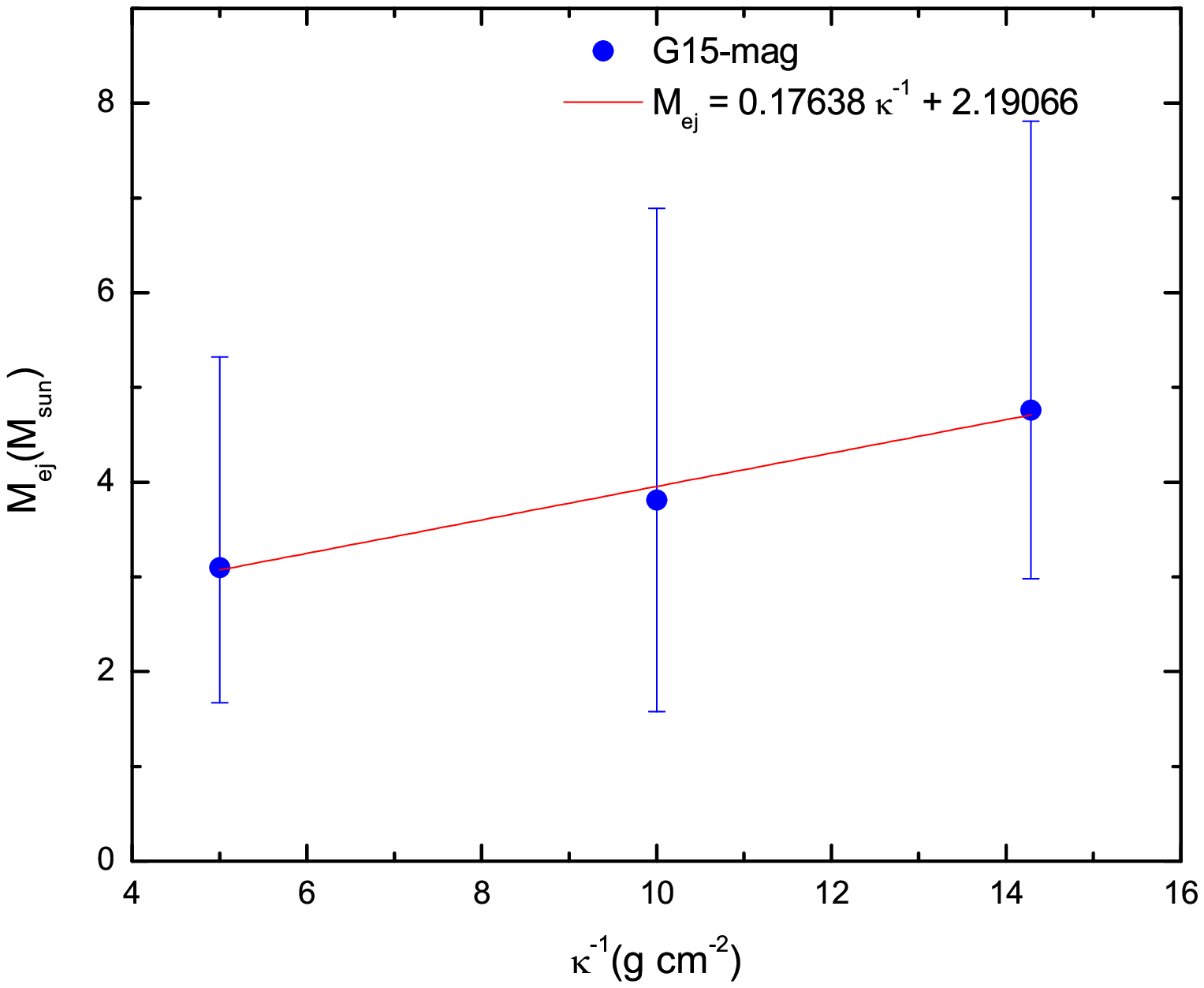}\\[0pt]
\includegraphics[width=0.45\textwidth,angle=0]{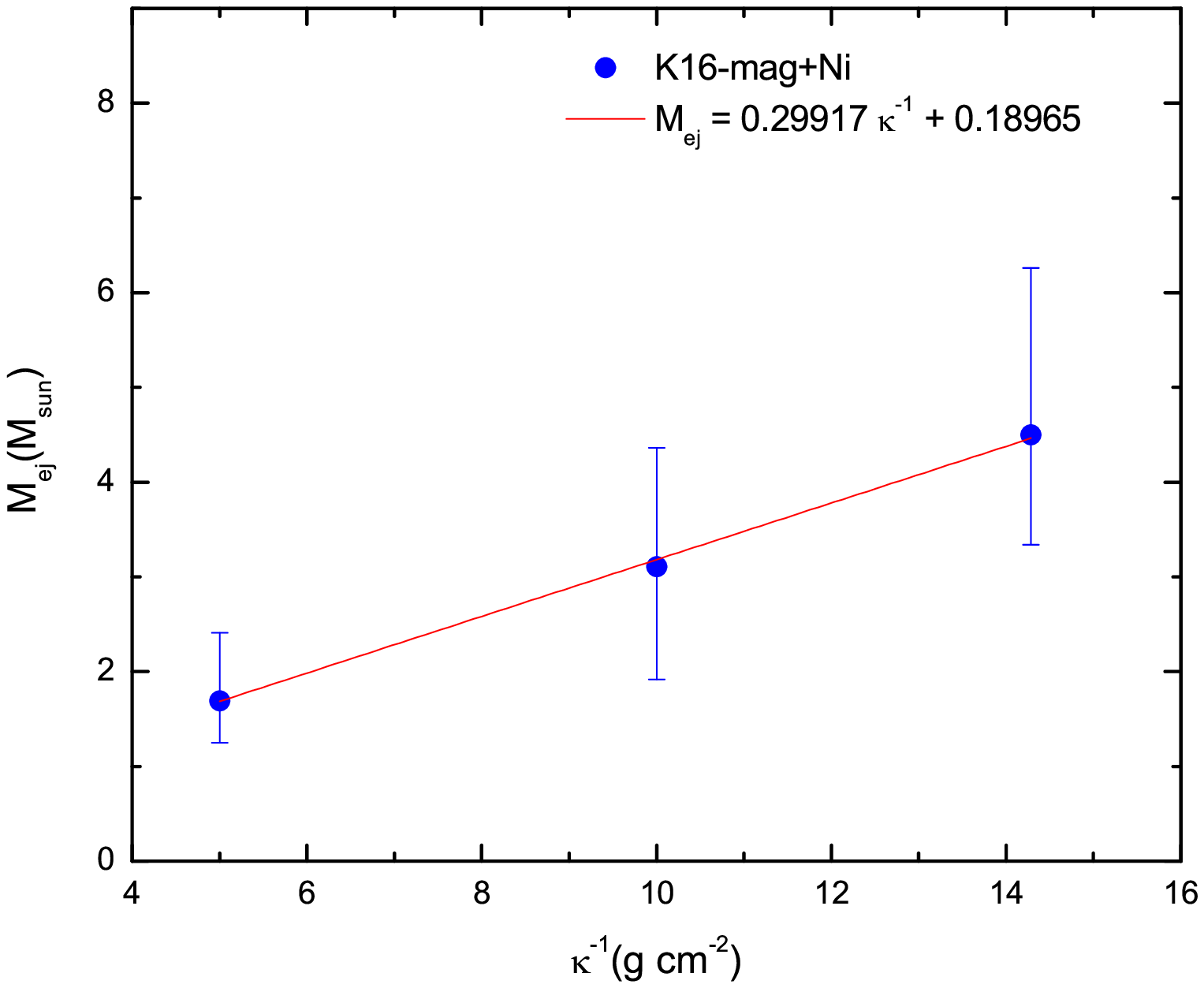}
\includegraphics[width=0.45\textwidth,angle=0]{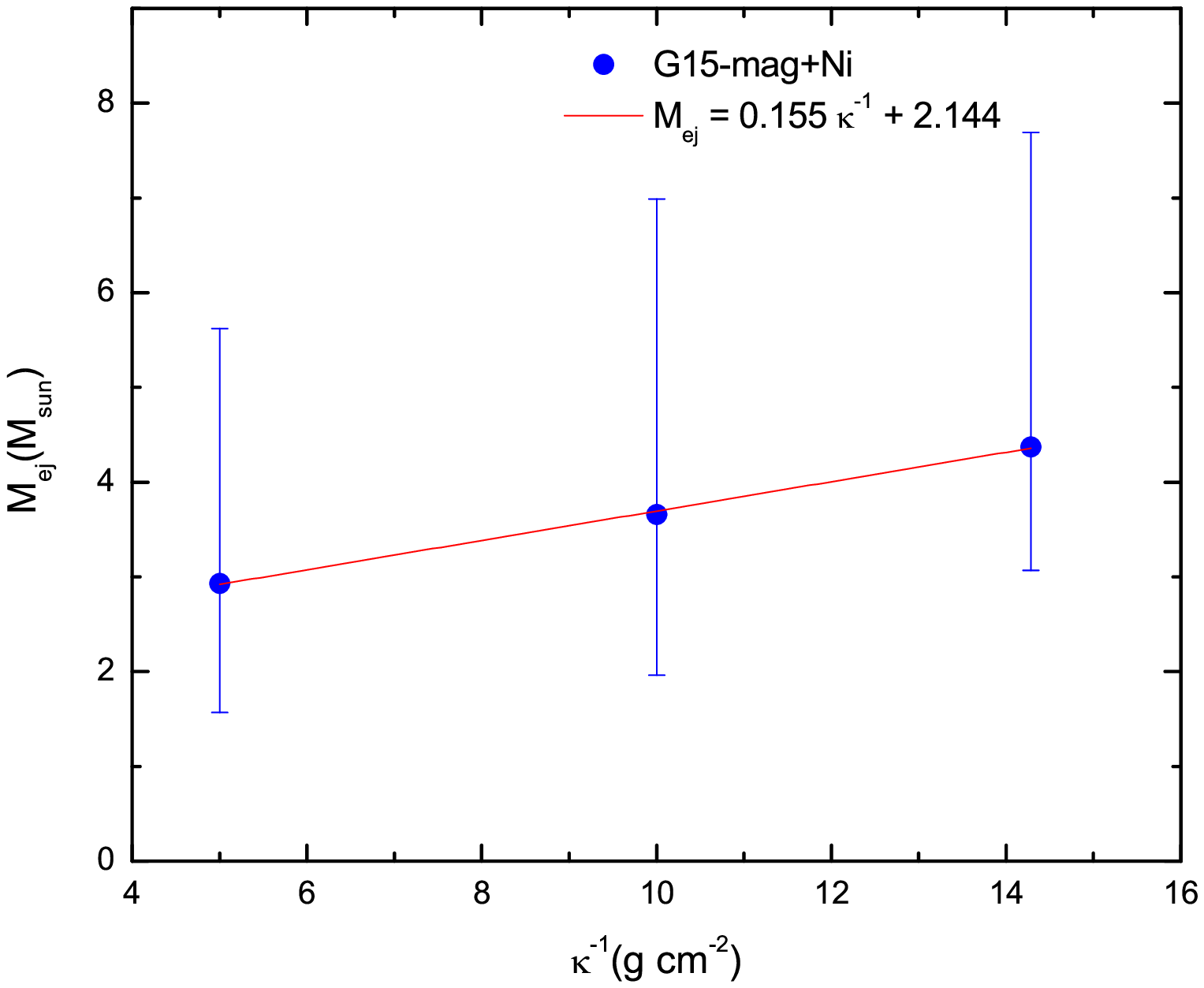}
\end{center}
\caption{Correlation between the ejecta mass and the optical opacity for SN~2011kl.}
\label{fig:cor}
\end{figure}

By comparing the values of the parameters corresponding to different
values for $\kappa$, we found that while the values of the ejecta masses are
significantly influenced by the values of $\kappa$ and $M_{\mathrm{ej}} =
a{\kappa}^{-1} +b$ ($a$ and $b$ are constants; see Figure \ref{fig:cor}, and a
similar figure can be found in \citealt{Nagy2016}), all other
parameters are slightly affected by the values of $\kappa$ and no
correlation between them and $\kappa$ has been found.

We did not study the correlations between the opacity and other
parameters for the I16 LC, since its data lack error bars, and the
value of $M_{\mathrm{ej}}$ is proportional to that of $\kappa ^{-1}$ while
the values of all other parameters are completely independent of that of
$M_{\mathrm{ej}}$.

\subsection{The Initial Kinetic Energy of SN~2011kl}

The multidimensional simulations (see \citealt{Jan2016}, and references
therein) for neutrino-driven SNe suggest that the value of $E_{\mathrm{K}}$
provided by the neutrinos can be up to $\sim$ (2.0--2.5) $\times 10^{51}$
erg. Provided the value of $v_{\mathrm{sc0}}$ is 14,000 km s$^{-1}$ (then
the ejecta mass is lowered to 2.4 M$_\odot$ for the K16~LC or 1.4 M$_\odot$
for the I16~LC), the origin of $E_{\mathrm{K,0}}$ of SN~2011kl
($E_{\mathrm{K,0}}=0.3\,M_{\mathrm{ej}}v_{\mathrm{sc0}}^{2} \approx 2.4\times
10^{51}$ erg) can be comfortably explained by the neutrino-driven model
and other processes are not required to provide additional $E_{\mathrm{K,0}}$.

According to the equation $E_{\mathrm{rot,0}}\approx (1/2)I\Omega_{0}^2
\approx 2\times 10^{52} \left({P_{0}}/{1~\mathrm{ms}}\right)^{-2}$~erg ($I
\approx 10^{45}{~\mathrm{g}~\text{\textrm{cm}}}^2$ is the rotational inertia
of the magnetar), we find that the initial rotational energy of the magnetar
($P_{0}=11.4$--14.2~ms) powering the LCs of SN~2011kl is $\sim 1.5 \times
10^{50}$~erg (\citealt{Cano2016} obtained $E_{\mathrm{rot,0}}=$ (1.2--1.6)
$\times10^{50}$~erg), significantly smaller than their $E_{\mathrm{K,0}}$.
Even if the initial rotational energy of this magnetar is all converted to
the $E_{\mathrm{K,0}}$ of the ejecta, the ratio of final $E_{\mathrm{K}}$ to
$E_{\mathrm{K,0}}$ is $\sim 1.1$. Therefore, the acceleration effect is
rather small and can be neglected.

\subsection{The Ejecta Mass and Binarity}

Assuming that $\kappa = 0.07$ cm$^2$ g$^{-1}$ and according to
the results of our work based on the magnetar models (the $^{56}$Ni
models have been excluded), the inferred ejecta mass of SN~2011kl
is $3.83_{-1.46}^{+2.06}$ M$_\odot$ (for the K16~LC), $
4.76_{-1.78}^{+3.05}$ M$_\odot$ (for the G15~LC), or $\sim 2.12$ M$_\odot$
(for the I16~LC). If $\kappa = 0.10$ cm$^2$ g$^{-1}$ or
$0.20$ cm$^2$ g$^{-1}$ were adopted, the inferred masses would decrease
by a factor of about 2--3, $\sim1$--2 M$_\odot$.

By adopting $\kappa = 0.04$ cm$^2$ g$^{-1}$ and $v$ ($v_{\mathrm{sc0}}$) =
20,000 km s$^{-1}$, \citet{Ioka2016} concluded that the mass of the ejecta
is 2 M$_\odot$ if SN~2011kl is powered by a magnetar. Assuming that $\kappa
= 0.07$ cm$^2$ g$^{-1}$ and $v$ ($v_{\mathrm{sc0}}$) = 21,000 km s$^{-1}$,
the ejecta mass must be 2 M$_\odot \times0.04/0.07 \times21,000/20,000
=1.323 $ M$_\odot$, significantly smaller than the value derived here (2.12
M$_\odot $).

This difference is caused by the fact that we adopted the equation
$\tau_{m}=(2{\kappa}M_{\mathrm{ej}}/{\beta}v_{\mathrm{sc}}c)^{1/2}$ while
\citet{Ioka2016} adopted the equation
$\tau_{m}=(3{\kappa}M_{\mathrm{ej}}/4{\pi}v_{\mathrm{sc}}c)^{1/2}$.
The ratio of the masses derived from these two
equations is $3\beta/8 {\pi}=3\times13.8/8\times3.1416=1.647$. By
multiplying this factor by 1.323 M$_\odot$, we obtain 2.18 M$_\odot$,
consistent with the value derived here (2.12 M$_\odot$). The ejecta mass
would be 1.45 M$_\odot$ if $\kappa = 0.1$ cm$^2$ g$^{-1}$ and $\tau_{m}=
(2{\kappa}M_{\mathrm{ej}}/{\beta}v_{\mathrm{sc}}c)^{1/2}$.

\citet{Ioka2016} and \citet{Arc2016} argued that the ejecta mass of
SN~2011kl inferred from the magnetar model is too low to be produced by a
core-collapse SN whose remnant harbors a magnetar. However, it must be
pointed out that such a small value is not disfavored by the explosion of a
SN~Ic and magnetar formation, since the massive hydrogen and helium
envelopes of the progenitors of SNe~Ic had been stripped by their companions
or winds, and the mean value of the ejecta mass of SNe~Ic is 2.9 M$_\odot$
with a standard deviation of 2.2 M$_\odot$ (see Table 6 of \citealt{Lym2016}). The inferred ejecta masses of Type Ic-BL SNe 2002ap, 2006aj, 2009bb, and
2010bh are $2.0_{-0.7}^{+0.8}$ M$_\odot $, $1.4_{-0.2}^{+0.4}$ M$_\odot$,
$1.9_{-0.5}^{+0.6}$ M$_\odot$, and $0.9_{-0.2}^{+0.2}$ M$_\odot$,
respectively (see Table 5 of \citealt{Lym2016}). Note, especially, that SN
2006aj with $M_{\mathrm{ej}}=1.4_{-0.2}^{+0.4}$ M$_\odot$ (or $\sim 2$
M$_\odot$; \citealt{Maz2006}) is a SN~Ic-BL associated with an X-ray flash (a
``soft GRB") that has long been believed to be powered by a nascent magnetar
\citep{Maz2006}.

Such small ejecta masses indicate that the progenitor of SN~2011kl might be
located in a binary system and most of its mass had been stripped by its
companion before the explosion. Although the mass-transfer process reduced
the mass of the progenitor, it can enhance the angular momentum and angular
velocity of the progenitor, making it easier for the nascent neutron stars
left by the SN explosions to be millisecond ($P_0 \approx 1$--10~ms)
magnetars.

\section{Conclusions}

\label{sec:con}

In this paper, we modeled the LCs of SN~2011kl, which is the most luminous (
$L_{\mathrm{peak}}~=~3.63^{+0.17}_{-0.16}\times 10^{43}$~erg~s$^{-1}$)
GRB-SN identified thus far. By using the bolometric LC of SN~2011kl derived
by \citet{Kann2016} and assuming that this SN is powered by $^{56}$Ni
cascade decay, we find that the required $^{56}$Ni mass is $
1.42_{-0.04}^{+0.04}$ M$_\odot$, which is smaller than that ($2.27\pm0.64$
M$_\odot$) inferred by \citet{Kann2016}. The $^{56}$Ni model can be excluded
from explaining the LCs of SN~2011kl, since the spectral features indicate
that the amount of $^{56}$Ni must be significantly smaller than $\sim 1$
M$_\odot$ \citep{Gre2015}.

Alternatively, we used the magnetar model and the magnetar+$^{56}$Ni model
to fit the K16~LC and the G15~LC, finding that both of the models can well
explain these LCs. It is noteworthy that \citet{Ioka2016} argue that the
fact that the location of GRB 111209A-SN~2011kl is near the nucleus of the
host galaxy favors the WD-TDE model. However, the position of the SN does
not necessarily mean that this event must be a TDE rather than a SN. The
magnetar model is still possible, explaining the K16~LC and the G15~LC,
while the TDE model is also a plausible one for the G15~LC and the K16~LC.

Whereas the magnetar model and the magnetar+$^{56}$Ni model can account for
the K16~LC and the G15~LC, they cannot explain the I16~LC because these
models cannot produce the early-time excess and the dip, although the
presence of the early-time excess is entirely model-dependent (i.e., the
parameters of the optical-afterglow component in the LC decomposition
technique). To solve this problem, we added the contribution of the cooling
emission from a low-mass ($M_{\mathrm{env}}\approx 0.5$ M$_\odot$), extended
($R_{\mathrm{env}}\approx 50$--100 R$_\odot$) SN progenitor envelope, and
found that the magnetar+cooling model and the magnetar+$^{56}$Ni+cooling
model can account for the I16~LC. In these models, cooling emission from the
shock-heated envelope powers the early-time excess, while the magnetar or
magnetar+$^{56}$Ni power the main peak of the LC, and the sum of these two
produces the dip.\footnote{It should
be mentioned that the model in which the progenitors have a
nonstandard structure \citep{NP2014,Piro2015} has been employed to explain
some SLSNe whose bolometric LCs exhibit an early-time excess
\citep{Smit2016,Vre2017}.} Hence, we conclude that the BSG model is not
unique for explaining the I16~LC \emph{even if} the LC of SN~2011kl has an
early-time excess, since a progenitor surrounded by a low-mass, extended
envelope can also power a LC with an early excess.

It seems that the magnetar+$^{56}$Ni+cooling and the magnetar+$^{56}$Ni
models are more reasonable than the magnetar and magnetar+cooling models for
the LCs of SN~2011kl, since core-collapse SNe must synthesize a moderate
amount of $^{56}$Ni. However, discriminating between the LCs reproduced by
the models with and without $^{56}$Ni is very difficult because the
contribution of a moderate amount of $^{56}$Ni ($\sim 0.1-0.2$ M$_\odot$) is
significantly smaller than that of the magnetar.

Provided that the initial velocity of the ejecta of SN~2011kl is $\sim
14,000 $ km s$^{-1}$ (the lower limit of the ejecta velocity; see
\citealt{Kann2016}), the inferred values of the initial kinetic energy of
this SN is $E_{\mathrm{K,0}} \approx 2.0 \times 10^{51}$~erg , indicating
that the neutrino-driven mechanism \citep{Jan2016} is able to provide the
$E_{\mathrm{K,0}}$ for this SN. But larger velocities require other
mechanisms to provide additional $E_{\mathrm{K,0}}$.

Furthermore, we used a MCMC method for the G15 and the K16 LCs to
constrain the range of the model parameters (we did not perform MCMC for
the I16 LC owing to the absence of the error bars); see
Table \ref{tab:para}.
By adopting different values of $\kappa$, we found that while
the inferred mass is significantly influenced by the values
of $\kappa$ ($M_{\mathrm{ej}} = a{\kappa}^{-1} +b$ ($a$ and $b$ are
constants), all other parameters are only slightly affected by the values
of $\kappa$ and no correlation between them and $\kappa$ has been found.

According to these results, we suggest that the magnetar and the
magnetar+$^{56}$Ni models,
with or without the cooling effect, can reproduce the LCs
of SN~2011kl.\footnote{We caution that the question
of how the magnetar powers an ultralong GRB is
still unsolved. \citet{Met2015} suggested that a magnetar with $P_0 \approx
2 $~ms and $B_{p} \approx 4\times10^{14}$~G can power the ultralong
GRB~111209A and SN~2011kl, but \citet{Beni2017} demonstrated that it is
difficult for a magnetar to produce an ultralong GRB. This issue is beyond
the scope of this paper and needs additional research. Moreover, we do not
attribute the optical afterglow of GRB~111209A to magnetar spin-down
emission.} In other words, SN~2011kl might be primarily powered by a nascent
magnetar.

\acknowledgments We thank I. Arcavi for helpful discussions of the cooling
model. This work is supported by the National Basic Research Program
(``973" Program) of China (grant 2014CB845800) and the
National Natural Science Foundation of China (grant 11573014). L.J.W. is
supported by the National Program on Key Research and Development Project of
China (grant 2016YFA0400801). A.V.F.'s supernova group is supported by
the Christopher R. Redlich Fund, the TABASGO Foundation, and the Miller
Institute for Basic Research in Science (U.C. Berkeley).


\clearpage

\clearpage

\clearpage

\clearpage

\begin{table*}[tbp]
\caption{Parameters of various models for SN~2011kl. The uncertainties are 1$\sigma$.}
\label{tab:para}
\par
\begin{center}
{\tiny
\begin{tabular}{ccccccccccccc}
\hline\hline
& $\kappa$ & $M_{\mathrm{ej}}$ & $M_{\mathrm{env}}$ & $R_{\mathrm{env}}$ &
$M_{\mathrm{Ni}}$ & $B_p$ & $P_0$ & $v_{\mathrm{sc0}}$ & $\kappa_{\gamma,
\mathrm{Ni}}$ & $\kappa_{\gamma,\mathrm{mag}}$ & $\chi^2$/dof &  \\
& (cm$^2$ g$^{-1}$) & (M$_\odot$) & (M$_\odot$) & (R$_\odot$) & (M$_\odot$)
& ($10^{14}$~G) & (ms) & (km~s$^{-1}$) & (cm$^2$ g$^{-1}$) & (cm$^2$ g$^{-1}$) &  &  \\ \hline\hline
\textbf{$^{56}$Ni} &  &  &  &  &  &  &  &  &  &  &  &  \\
&  &  &  &  &  &  &  &  &  &  &  &  \\ \hline
A1(K16ni007) & 0.07 & $4.57_{-1.03}^{+0.80}$ & - & - & $1.42_{-0.04}^{+0.04}$
& - & - & $22,905.4_{-4,924.7}^{+3,546.1}$ & 0.027 & - & 0.353 &  \\
A1$^{\prime}$(K16ni010) & 0.10 & $3.24_{-0.78}^{+0.74}$ & - & - & $1.48_{-0.04}^{+0.04}$ & - & - & $21,710.9_{-4,636.7}^{+4,143.7}$ & 0.027 & -
& 0.356 &  \\
A1$^{\prime\prime}$(K16ni020) & 0.20 & $1.71_{-0.40}^{+0.56}$ & - & - &
$1.64_{-0.05}^{+0.06}$ & - & - & $19,535.3_{-3,827.8}^{+4,954.4}$ & 0.027 & -
& 0.360 &  \\
A2(G15ni007) & 0.07 & $3.32_{-0.79}^{+0.94}$ & - & - & $1.06_{-0.05}^{+0.05}$
& - & - & $20,387.7_{-4,341.8}^{+4,867}$ & 0.027 & - & 0.061 &  \\
A2$^{\prime}$(G15ni010) & 0.10 & $2.44_{-0.59}^{+0.77}$ & - & - & $1.11_{-0.05}^{+0.06}$ & - & - & $19,868.6_{-4,066.9}^{+4,999.6}$ & 0.027 & -
& 0.062 &  \\
A2$^{\prime\prime}$(G15ni020) & 0.20 & $1.50_{-0.35}^{+0.48}$ & - & - & $1.25_{-0.06}^{+0.07}$ & - & - & $19,993.7_{-3,723.3}^{+4,941.9}$ & 0.027 & -
& 0.092 &  \\
A3(I16ni007) & 0.07 & 2.12 & - & - & 1.1 & - & - & 21,000 & 0.027 & - & - &
\\ \hline\hline
\textbf{magnetar} &  &  &  &  &  &  &  &  &  &  &  &  \\
&  &  &  &  &  &  &  &  &  &  &  &  \\ \hline
B1(K16mag007) & 0.07 & $3.83_{-1.46}^{+2.06}$ & - & - & 0 & $5.99_{-5.55}^{+1.75}$ & $10.83_{-6.94}^{+0.49}$ & $21,715.3_{-4,973.8}^{+4,293.1}$ & - & $1,918.79_{-1,918.775}^{+5,486.67}$ &
0.259 &  \\
B1$^{\prime}$(K16mag010) & 0.10 & $3.20_{-0.73}^{+1.23}$ & - & - & 0 & $6.62_{-2.15}^{+1.37}$ & $10.94_{-3.35}^{+0.38}$ & $21,159.4_{-4,045.1}^{+4,768.8}$ & - & $4,426.31_{-3,564.503}^{+3,728.79}$ &
0.398 &  \\
B1$^{\prime\prime}$(K16mag020) & 0.20 & $1.94_{-0.58}^{+1.16}$ & - & - & 0 &
$6.75_{-1.38}^{+1.35}$ & $11.08_{-0.61}^{+0.30}$ & $20,672.7_{-4,609.8}^{+4,445.9}$ & - & $2,781.08_{-2,781.0756}^{+4,888.58}$ &
0.361 &  \\
B2(G15mag007) & 0.07 & $4.76_{-1.78}^{+3.05}$ & - & - & 0 & $9.72_{-4.13}^{+3.23}$ & $12.07_{-2.52}^{+1.15}$ & $20,846.3_{-4,860.9}^{+3,887.4}$ & - & $4,478.71_{-3,969.36}^{+3,724.93}$ &
0.065 &  \\
B2$^{\prime}$(G15mag010) & 0.10 & $3.81_{-2.23}^{+3.08}$ & - & - & 0 & $9.40_{-8.14}^{+4.68}$ & $10.01_{-4.74}^{+2.89}$ & $21,517.1_{-4,610.3}^{+4,308.8}$ & - & $2,559.96_{-2,559.941}^{+4,995.82}$ &
0.066 &  \\
B2$^{\prime\prime}$(G15mag020) & 0.20 & $3.10_{-1.43}^{+2.22}$ & - & - & 0 &
$12.20_{-3.97}^{+4.80}$ & $9.26_{-3.35}^{+3.41}$ & $20,798.6_{-5,027.5}^{+5,404.0}$ & - & $3,855.41_{-3,764.804}^{+4,101.43}$ &
0.066 &  \\
B3(I16mag007) & 0.07 & 2.12 & - & - & 0 & 6.5 & 13.2 & 21,000 & - & $10^4$ &
- &  \\ \hline\hline
\textbf{magnetar+$^{56}$Ni} &  &  &  &  &  &  &  &  &  &  &  &  \\
&  &  &  &  &  &  &  &  &  &  &  &  \\ \hline
C1(K16magni007) & 0.07 & $4.50_{-1.16}^{+1.76}$ & - & - & $0.11_{-0.07}^{+0.06}$ & $6.94_{-1.62}^{+1.51}$ & $11.46_{-0.79}^{+0.46}$ & $21,763.4_{-4,977.1}^{+4,440.2}$ & 0.027 & $4,229.65_{-3,633.31}^{+3,885.96}$
& 0.409 &  \\
C1$^{\prime}$(K16magni010) & 0.10 & $3.11_{-1.19}^{+1.25}$ & - & - & $0.11_{-0.06}^{+0.06}$ & $6.80_{-1.85}^{+1.58}$ & $11.3_{-1.01}^{+0.46}$ & $20,334.9_{-4,638.9}^{+5,272.6}$ & 0.027 & $4,040.85_{-3,776.55}^{+3,948.94}$
& 0.413 &  \\
C1$^{\prime\prime}$(K16magni020) & 0.20 & $1.69_{-0.44}^{+0.72}$ & - & - & $0.11_{-0.07}^{+0.05}$ & $7.12_{-1.43}^{+1.28}$ & $11.41_{-0.93}^{+0.41}$ & $21,537.6_{-4,485.8}^{+5,034.7}$ & 0.027 & $3993.19_{-3975.58}^{+4,038.74}$ &
0.406 &  \\
C2(G15magni007) & 0.07 & $4.37_{-1.30}^{+3.32}$ & - & - & $0.10_{-0.07}^{+0.06}$ & $10.07_{-7.30}^{+3.93}$ & $12.21_{-3.60}^{+1.65}$ & $19,410.8_{-4,122.9}^{+5,156.7}$ & 0.027 & $3,670.59_{-3,670.57}^{+4,243.63}$
& 0.071 &  \\
C2$^{\prime}$(G15magni010) & 0.10 & $3.66_{-1.70}^{+3.33}$ & - & - & $0.11_{-0.07}^{+0.05}$ & $10.48_{-9.10}^{+5.04}$ & $10.80_{-5.95}^{+2.83}$ & $21,051.0_{-4,415.6}^{+4,496.1}$ & 0.027 & $2,767.03_{-2,766.996}^{+4,768.41}$
& 0.071 &  \\
C2$^{\prime\prime}$(G15magni020) & 0.20 & $2.93_{-1.36}^{+2.69}$ & - & - & $0.10_{-0.07}^{+0.07}$ & $14.10_{-4.33}^{+5.12}$ & $10.47_{-5.62}^{+2.87}$ & $20,857.9_{-4,538.1}^{+4,792.0}$ & 0.027 & $5007.93_{-3376.97}^{+3334.9}$ &
0.069 &  \\
C3(I16magni007) & 0.07 & 2.12 & - & - & 0.2 & 6.5 & 14.2 & 21,000 & 0.027 &
0.13 & - &  \\ \hline\hline
\textbf{magnetar+$^{56}$Ni} &  &  &  &  &  &  &  &  &  &  &  &  \\
\textbf{+cooling} &  &  &  &  &  &  &  &  &  &  &  &  \\ \hline
D1 (I16cooling) & 0.07 & 2.12 & 0.63 & 51.4 & 0 & - & - & 21,000 & - & - & -
&  \\
D2(I16magnicooling)$^{a}$ & 0.07 & 2.12 & 0.63 & 51.4 & 0.2 & 6.5 & 14.2 &
21,000 & 0.027 & 0.13 & - &  \\
D2$^{\prime}$(I16magnicooling)$^{a}$ & 0.07 & 1.6 & 0.45 & 103 & 0.2 & 6.5 &
14.2 & 16,000 & 0.027 & 0.13 & - &  \\ \hline\hline
\end{tabular}
}
\end{center}
\par
{\tiny
$^{a}$The cooling+magnetar+$^{56}$Ni model (D2 = D1 + C3). \newline
}
\end{table*}

\clearpage

\end{document}